\documentclass{article}
\usepackage{wasysym}
\usepackage{graphicx}
\usepackage[usenames]{color}
\usepackage{itmath}
\usepackage{textcomp}
\usepackage{my-latex}
\newcommand{\CITSOV}[0]{\mathsf{CITSOV}}
\newcommand{\NODICT}[0]{\mathsf{NODICT}}
\newcommand{\BR}[0]{\mathsf{BR}}
\newcommand{\DOM}[0]{\mathsf{DOM}}
\newcommand{\MON}[0]{\mathsf{MON}}
\newcommand{\STRPROOF}[0]{\mathsf{STRPROOF}}
\newcommand{\ballot}[0]{\mathsf{ballot}}
\newcommand{\trueprofile}[0]{\mathsf{true}}
\newcommand{\considers}[3]{p^{#1}_{#2>#3}}
\newcommand{\putaway}[1]{}

  \title{Reasoning about Social Choice Functions}
\date{}
\author{Nicolas Troquard \and Wiebe van der Hoek \and Michael
  Wooldridge} \date{\small Computer Science Department, University of
  Liverpool, UK}

\begin{document} 
\maketitle

\begin{abstract}
  We introduce a logic specifically designed to support reasoning
  about social choice functions. The logic includes operators to
  capture strategic ability, and operators to capture agent
  preferences. We establish a correspondence between formulae in the
  logic and properties of social choice functions, and show that the
  logic is expressively complete with respect to social choice
  functions, i.e., that every social choice function can be
  characterised as a formula of the logic. We prove that the logic is
  decidable, and give a complete axiomatization. To demonstrate the
  value of the logic, we show in particular how it can be applied to
  the problem of determining whether a social choice function is
  strategy-proof.
\end{abstract} 

%%%%%%%%%%%%%%%%
%%% INTRODUCTION
\section{Introduction}

\emph{Social choice theory} is concerned with collective decision
making in situations where the preferences of the decision makers may
differ~\cite{handbook02SocChWelf}. Social choice theorists have
developed a range of procedures, such as voting protocols, to support
such collective decision making, and have developed a range of
criteria with which to characterise the properties of such
procedures. Such criteria are usually expressed axiomatically, and a
major concern of social choice theory is to study the extent to which
decision making procedures do or do not satisfy these
axioms~\cite{may52econometrica,arrow50,gibbard73,satterthwaite75}.

In short, the aim of the present paper is to develop a logic that is
explicitly intended for reasoning about social choice procedures. We
focus on \emph{social choice functions}, a class of social choice
procedures that select a single social outcome as a function of
individual preferences. \emph{Voting procedures} of the type used in
political elections throughout the democratic world are perhaps the
best-known examples of social choice functions. A voting procedure
determines the winner of an election as a function of the votes cast;
votes can be understood as an expression of voter preferences. 

One interesting issue that arises in voting procedures is the extent
to which voters are incentivised to truthfully report their
preferences when voting.  For example, suppose we have two voters, $1$
and $2$, who vote among three candidates, $x$, $y$, and $z$ for a role
that is currently filled by $x$. The voting procedure used in this
example says that, if there is a unanimously preferred candidate, then
that will be chosen, otherwise the candidate $x$ remains. Suppose the
true preferences of $1$ are given by $z <_1 x <_1 y$ and those of $2$
are $x <_2 y <_2 z$. If the social choice function was presented with
these true preferences, candidate $x$ would be chosen (since there is
no consensus). However, if voter $2$ would instead claim his
preferences were $x <'_2 z <'_2 y$ while $1$ revealed its true
preferences, then $2$ would be better off, since $y$ would be chosen,
rather than $x$, and agent $2$ prefers $y$ over $x$.  This issue
suggests the following problem: Can we design a voting procedure that
is ``immune'' to such misrepresentation, i.e., in which a voter can
never do any better than by truthfully reporting its preferences? The
term \emph{strategy proof} is used to refer to such voting procedures.
In fact, fundamental results in social choice theory tell us that
there are severe limits to the development of strategy-proof voting
procedures~\cite{gibbard73,satterthwaite75}, and for this reason,
developing and analysing social choice procedures is a lively and
highly active research area.

The long-term aim of our work is to develop formal tools to assist in
the analysis and design of social choice procedures. In particular, we
hope to develop techniques that will permit the \emph{automated}
analysis of social choice procedures. To this end, we aim to develop
logics that allow us to formally express the properties of social
choice procedures, such that these languages may be automatically
processed.  Our view is that logic can provide a powerful tool for the
analysis of social choice
procedures~\cite{pauly01thesis,wooldridge:2007a}. Such logics can be
used as \emph{query languages} for social choice procedures: given
some property $P$ of a social choice procedure, we aim to be able to
encode the property $P$ as an expression $\rho^P$ of our language,
which we then pose as a query to an automated analysis system.
Working towards the long-term goal, the present paper presents a logic
for reasoning about social choice procedures, and in particular, for
analysing strategy proofness.  

The remainder of the paper is structured as follows.  In
Section~\ref{sec:background} we recall the main concepts from game
theory and social choice theory that we use throughout the paper.  We
then introduce our logic in Section~\ref{sec:lscf}.  The logic is
basically a modal logic \cite{ChellasML}, which derives inspiration
from the Coalition Logic of Propositional Control (\CLPC)
\cite{HoekWooldridge:AIJ2005}. The latter logic includes operators to
capture strategic ability. We extend this with operators for capturing
agent preferences.  The basic idea is to model an agent's preferences
via atomic propositions: a proposition $\considers{i}{x}{y}$ will be
used to represent the fact that agent $i$ has reported that he prefers
outcome $x$ at least as much as outcome $y$. The strategic abilities
of agents are captured using a \CLPC-like operator: an agent can
choose any assignment of values for its preference variables that
corresponds to a preference ordering.  After presenting the syntax and
semantics of the logic, we show how the logic can be used to
characterise social choice functions, and show that the logic is
expressively complete with respect to social choice functions, i.e.,
that every social choice function can be characterised as a formula of
the logic. We give a complete axiomatization for the logic. To
demonstrate the value of the logic, in Section~\ref{sec:applications}
we formalise some properties of social choice functions and in
particular, we show how it can be applied to the problem of
determining whether a social choice function is strategy-proof. We
conclude in Section~\ref{sec:conclusion}.

% 

%%%%%%%%%%%%%
%%% BACGROUND
\section{Background}
\label{sec:background}

In this section, we present the basic definitions of game theory and
social choice upon which we construct our
framework~\cite{dasgupta79inc-comp,osborne94}.
% for talking about social choice functions.

We begin with some notation. We assume that game forms and social
choice functions (to be defined hereafter) share the same domains of
agents and outcomes. We denote by $N = \{1, \dots, n\}$ the finite set
of agents (or players) and by $K$ the finite set of social outcomes
(outcomes hereafter). We use the letters $a, b, c, \dots$ as constants
of $K$. We use variables $i, j, \dots$ to denote agents, and outcomes
will be denoted by the variables $x, y, z, \dots$.  Typically, one can
consider that the agents are the voters and the outcomes are the
candidates in some election.

We denote by $L(K)$ the set of \emph{linear} orders over $K$. (A
linear order here is a relation that is reflexive, transitive,
antisymmetric and total.)  By using a linear order, we are assuming
the players cannot be indifferent between two \emph{distinct}
outcomes. A \emph{preference relation} is a linear order of
outcomes. Given $K$ and $N$, a \emph{preference profile} $<$ is a
tuple $(<_i)_{i \in N}$ of preferences, where $<_i\ \in L(K)$ for
every $i$. The set of preference profiles is denoted by $L(K)^N$. Note
that we use the symbol $<_i$ for a preference relation for agents,
which in this case happens to be reflexive (and we do not write
$\leq_i$ for it). Also, we will use the symbol $>_i$ with the obvious
meaning, i.e., $y >_i x$ iff $x <_i y$.

\begin{definition}[Social choice function]
  Given $K$ and $N$, a \emph{social choice function} (SCF) is a
  single-valued mapping from the set $L(K)^N$ of preference profiles
  into the set $K$ of outcomes.
\end{definition}
For every preference profile, a social choice function describes the
desirable outcome (from the point of view of the designer).

\begin{definition}[Strategic game form]
  Given the sets $N$ and $K$, a \emph{strategic game form} is a tuple $\tuple{N, (A_i), K, o}$
  where:
  \begin{itemize*}
 % \item $N$ is a finite nonempty set of \emph{players} (or agents);
  \item[] $A_i$ is a finite nonempty set of \emph{actions} (or
    strategies) for each player $i \in N$;
  %\item $K$ is a finite nonempty set of outcomes;
  \item[] $o: \times_{i \in N} A_i \rightarrow K$ assigns an outcome
    for every combination of actions.
  \end{itemize*}
\end{definition}
A strategic game form is sometimes called a \emph{mechanism}. It
specifies the agents taking part in the game, their available actions,
and what outcome results from each combination of actions.
We refer to a collection $(a_i)_{i \in N}$, consisting of one action
for every agent in $N$, as an \emph{action profile}.  Given an action
profile $a$, we denote by $a_i$ the action of the player $i$.

%, and by $a_{-i}$ the collective action of the coalition $N \setminus
%\{ i \}$. For any $C \subseteq N$, we write $A_C$ for the
%\emph{coalitional actions} of those players that are a member of $C$,
%i.e., $A_C = \times_{j \in C} A_j$.

\begin{remark}\label{rem:moulin}
  There is a direct link between strategic game forms and social
  choice functions. Any social choice function can be viewed as a game
  form in which the set of actions of every agent is $L(K)$ (think of
  this as the preference profiles the agent can claim to be his), and
  the function $o$ represents the social choice function (see
  \cite{moulin83strategy}). For any SCF $F$, we denote its associated
  game form by $g^F$.
\end{remark}

%%\medskip\noindent\textbf{Strategic games and solution concepts.}~

A \emph{strategic game} is basically the composition of a strategic
game form with a collection of preference
relations (one for every agent) over the set of outcomes.
\begin{definition}[Strategic game]\label{def:str-game}
  A \emph{strategic game} is a tuple $\tuple{N, {(A_i)}, K,%\linebreak
    o, {(<_i)}}$ where $\tuple{N, (A_i), K, o}$ is a strategic game
  form, and for each player $i \in N$, $<_i$ is a \emph{preference
    relation} over $K$.
\end{definition}

In our context, when the actions $A_i$ in a game $\tuple{N, {(A_i)},
  K,%\linebreak 
o, {(<_i)}}$ are preference relations themselves, one
should think of those as preferences that $i$ can choose to report,
whereas $<_i$, encodes $i$'s real preferences.
%We refer to a collection $(a_i)_{i \in N}$, consisting of one action
%for every agent in $N$, as an \emph{action profile}.  Given an action
%profile $a$, we denote by $a_i$ the action of the player $i$, and by
%$a_{-i}$ the action profile of the coalition $N \setminus \{ i \}$. We
%write $a_C$ for the \emph{coalitional actions} that are members of
%$A_C = \times_{j \in C} A_j$ for any $C \subseteq N$.  We refer to a
%collection $(<_i)_{i \in N}$ of preferences as a
%\emph{preference profile}.

%%\medskip

A solution concept defines for every game a set of action profiles --
intuitively, those that may be played through rational action. Exactly
which solution concept is used depends upon the application at hand:
we will soon introduce a well-celebrated solution concept of {\em Nash
  Equilibrium} (see Example~\ref{ex:first}).

\begin{definition}[Solution concept]
  A \emph{solution concept} SC is a function that maps a strategic game
  form $\tuple{N, (A_i), K, o}$ and a preference profile over $K$ to a
  subset of the action profiles.
\end{definition}
We now introduce a simple but fundamentally important 
solution concept: Nash equilibrium.
\begin{definition}[Nash equilibrium]
Given a strategic game form %\linebreak 
$g =
  \tuple{N, (A_i), K, o}$ and a preference profile $<$ over
  $K$ the set of Nash equilibria $NE(g,<)$ is given as the set of
  action profiles in $g$ such that no player would benefit from
  deviating unilaterally from his current action.
  More formally, $(a_1, \ldots a_n) \in NE(g,<)$ iff for every player
  $k$ and every $a'_k \in A_k$, we have $o(a_1, \ldots a'_k \ldots
  a_n) <_k o(a_1, \ldots a_k \ldots a_n)$.
\end{definition}  

%%\medskip\noindent\textbf{Implementations.}~

We can now introduce the notions of implementation and truthful
implementation.  The problem of implementation arises because a
planner does not know the true preference profile of the
players. Given a social choice function $F$ involving a set of players
$N$ and a set of outcomes $K$, the planner only knows that every
player $i \in N$ has some preference $<_i$, an element of $L(K)$.

We first define the case of (standard) \emph{implementation}.
Assuming a pattern of behaviour -- a solution concept $SC$ -- the role
of the planner is then to design a mechanism (or game form) $g$ such
that for every possible preference profile $<\ \in L(K)^N$, the
strategic game $\tuple{g,<}$ admits at least one $SC$-equilibrium,
and every $SC$-equilibrium leads to the outcome in $K$ which is
prescribed by the social choice function for the preference profile at
hand, that is, the value of $F(<)$.

\begin{figure}[ht]
\begin{center}
  \framebox[\linewidth]{ 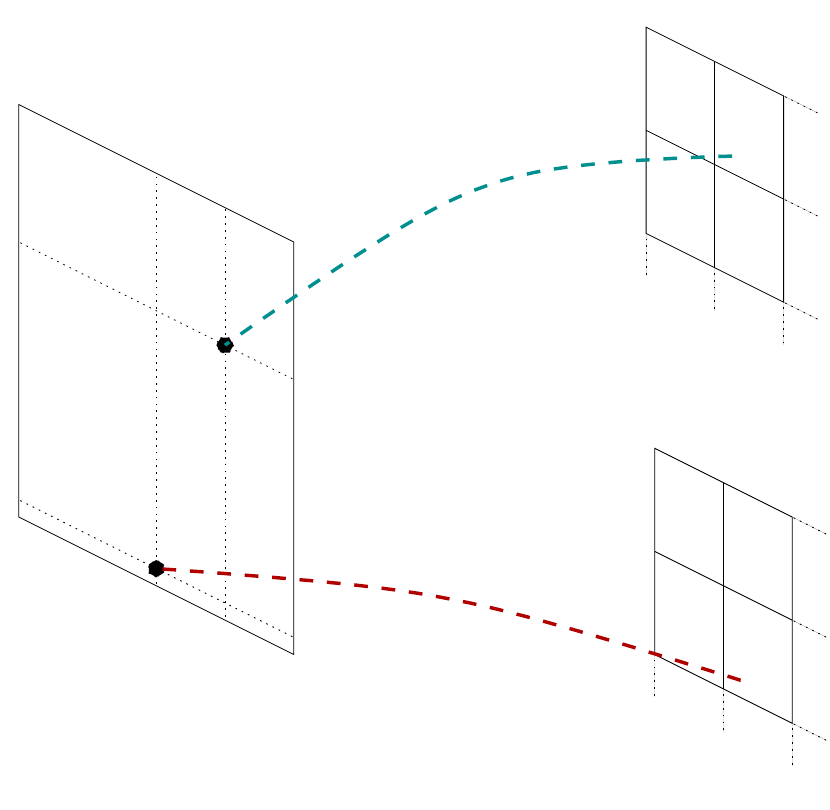 }
\end{center}
\caption{{\em Implementation.} 
  The preference profiles $<$
  and $<'$ are two arbitrary members of $L(K)^N$. The left part
  represents the SCF $F$.  $F(<_1,<_2) = y$ and $F(<'_1,<'_2) =
  x$. The right part represents the strategic game form $g$
  instantiated, in the upper part with the preference profile
  $(<_1,<_2)$ (game $G = \tuple{g, (<_1,<_2)}$) and in the lower part
  with the preference profile $(<'_1,<'_2)$ (game $G' = \tuple{g,
    (<'_1,<'_2)}$).
  All the $SC$-equilibria of $G$ (and possibly also some others than
  $(a'_1,a'_2)$) lead to $F(<_1,<_2)$. In a like manner, all the
  $SC$-equilibria of $G'$ lead to $F(<'_1, <'_2)$. This has to be
  verified for every preference profile in $L(K)^N$ and not only $<$
  and $<'$: if it holds, $g$ is said to $SC$-implement $F$.}
  \label{fig:implementation}
\end{figure}

\begin{definition}[Implementation]
 Given a solution concept $SC$, we say that the game form $g =
 \tuple{N, (A_i), K, o}$ \emph{$SC$-implements} the social choice
 function $F$ if for every preference profile $<\ \in L(K)^N$ we have
 that $SC(g,<) \not = \emptyset$ and
 \[a^* \in SC(g,<) \mbox{ implies that } o(a^*) = F(<) \]
\end{definition}
In words: the game form $g$ $SC$-implements $F$ if for any game form
$\tuple{g,<}$ based on $g$, any outcome associated to a strategy
profile in the solution concept $SC$ is the same as what the social
choice function would yield for the preference $<$. Or, more loosely:
the game form $g$ implements $F$ if, for every preference profile $<$
that we can associate with it, the outcomes in the game $\langle
g,<\rangle$ and the result of $F(<)$ agree at least on those claimed
preferences that are in the solution concept of the game.
 
The problem of implementation is illustrated in Figure
\ref{fig:implementation}. We say that the social choice function is
\emph{$SC$-implementable} if there is a game form that $SC$-implements
it.

%\medskip

In some situations however, an SCF can be implemented by a strategic
game form of which the space of action profiles corresponds to the
space of preference profiles, and \emph{telling the truth is an
  equilibrium}.  We call a strategic game form in which the set of
strategies of a player $i$ is the set of preferences over $K$ a
\emph{direct mechanism}. Hence, each player is asked to report a
preference, but not necessarily the true one.
An appealing class of direct mechanisms is that in which reporting the
true preference profile is an equilibrium of the game consisting of
the direct mechanism composed with the true preference profile. That
is, for every $< \in L(K)^N$, the action profile where every player
$i$ reports its true preference $<_i$ is an equilibrium of the game
$\tuple{g,<}$. We can define this notion for every solution concept
$SC$.

\begin{figure}[ht]
\begin{center}
  \framebox[\linewidth]{ 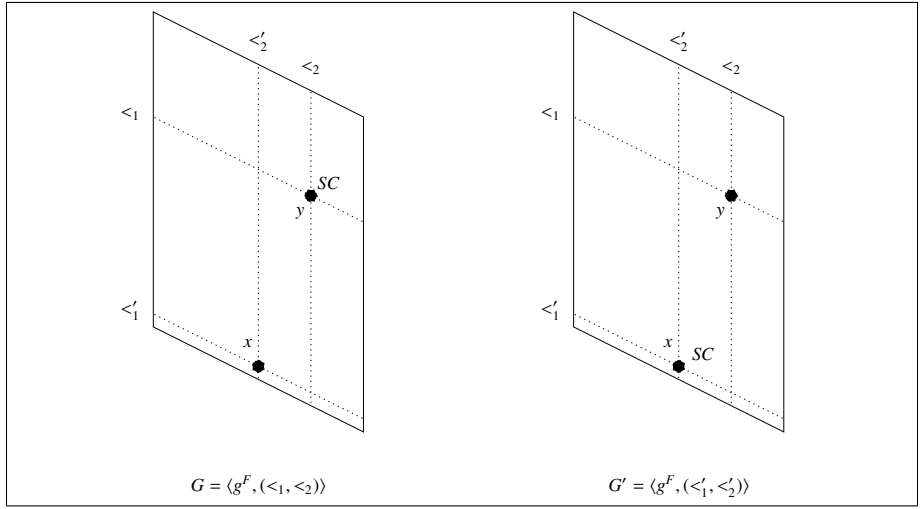 }
\end{center}
\caption{ 
  {\em Truthful implementation.} The preference profiles 
  $<$ and $<'$ are two arbitrary members of $L(K)^N$. The left 
  part represents the game form 
  $g^F$ associated to the SCF $F$ when the preferences of the two 
  players are $<_1$ and $<_2$. The game $G = \tuple{g^F, (<_1,<_2)}$ 
  admits an equilibrium at the action profile $(<_1,<_2)$. 
  The right part represents $g^F$ when the preferences of the two
  players are $<'_1$ and $<'_2$. The game $G' = \tuple{g^F,
    (<'_1,<'_2)}$ admits an equilibrium at the action profile
  $(<'_1,<'_2)$. This has to be verified for every preference profile
  in $L(K)^N$ and not only $<$ and $<'$: if it holds, $g^F$ is
  said to truthfully $SC$-implement $F$. }
  \label{fig:truth-implementation}
\end{figure}

\begin{definition}[Truthful implementation]
  The direct mechanism $g = \tuple{N,%\linebreak 
{(A_i)}, K, o}$ \linebreak\emph{truthfully
    $SC$-implements} the SCF $F$ if for every true preference profile $<$ and reported profile $a^*$ with $a^*_i = <_i$ for every $i$:
\[
a^* \in SC(g,<) \mbox{, and }
o(a^*) = F(<)
\]
\end{definition}

In words: $g$ is a truthful $SC$-implementation of $F$ if, for every profile $<$, whenever the agents declare that to be their real preferences, this a solution concept $SC$, and the outcome in the game and the function $F$ are the same. 
The problem of truthful implementation is illustrated on Figure
\ref{fig:truth-implementation}. We say that the social choice function
is \emph{truthfully $SC$-implementable} if there is a game form that
truthfully $SC$-implements it. Note that truthful implementations only
require that the report of the true preference profile is an
equilibrium, but it is not required that this equilibrium is
unique. In general, other equilibria could be present that would not
lead to the outcome prescribed by the SCF. However, this notion of
implementation can be motivated. Indeed, it is assumed that playing a
direct mechanism, if casting the real preference is an equilibrium
strategy, an agent would be sincere.

%%\medskip

We illustrate the differences between the problems of implementation
with some simple examples (a `minimal' social choice scenario with
only two voters and two alternatives), which demonstrates that the two
notions are contingent and independent: a game form $g$ can be both a
truthful $SC$-implementation and an $SC$-implementation of a social
function $F$, it can be both, and it can be either of them without
being the other.

  \begin{figure}%[h]
    \begin{center}
      \framebox[\linewidth] { 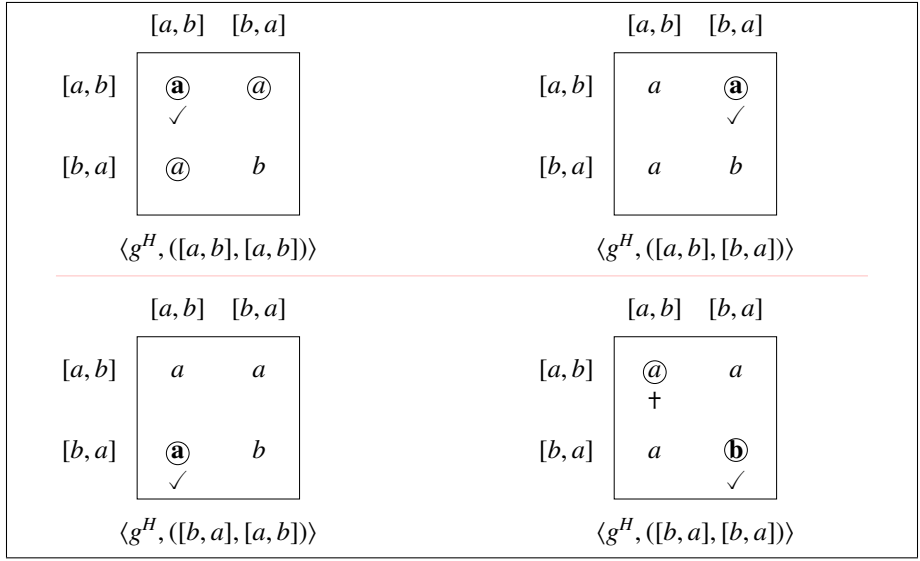 }
    \end{center}
    \caption{ $g^H$ does not $NE$-implement $H$. But $g^H$ truthfully
      $NE$-implements $H$.}
       \label{fig:ne-exampl}
  \end{figure} 
\begin{example}\label{ex:first}
  In this example we define some simple social choice functions, for
  all of them we set $N = \{1,2\}$ and $K = \{a,b\}$. Also, for the
  sake of comparison between standard and truthful implementations, we
  only consider direct mechanisms, since truthful implementations are
  not defined otherwise.
  
  First, consider the function $H$ for which we claim that its
  associated game form $g^H$ truthfully $NE$-implements $H$ but $g^H$
  does not $NE$-implement it.  $H$ is the social choice function
  prescribing the outcome $b$ if and only both agents prefer $b$ over
  $a$.  We write $[a,b]$ for the individual order of preferences of
  the outcome $a$ over the outcome $b$ and $[b,a]$ for the individual
  preference of $b$ over $a$. Hence, we have:
  \begin{itemize*}
  \item[] $H([a,b],[a,b]) = H([a,b],[b,a]) = H([b,a],[a,b]) = a$;
  \item[] $H([b,a],[b,a]) = b$.
  \end{itemize*}

  Figure \ref{fig:ne-exampl} represents the four possible games
  $\tuple{g^H,<}$ where $< \in L(\{a,b\})^{\{1,2\}}$. In each of them,
  the circles indicate the action profiles that are Nash
  equilibria. The outcomes in bold are the outcomes $o(a^*)$ for which
  $a^*\ =\ <$: in those outcomes, players have revealed their true
  preferences. So for instance, the outcome ${\mathbf a}$ in the upper
  left corner of the game $\tuple{g^H,([a,b],[a,b])}$ reads: `the
  outcome in the game here is $a$ and the voters reveal their true
  preferences'. For every preference profile $<$, the ticks
  \checked\ indicate that the action profile $<$ leads to the outcome
  prescribed by the social choice function $H$ and is a Nash
  equilibrium in the game $\tuple{g^H,<}$; Hence the game form $g^H$
  truthfully $NE$-implements $H$: all the bold outcomes are
  ticked. The cross \textdied\ designates a problem with the
  (standard) implementation of $H$ by $g^H$: in the game $\tuple{g^H,
    ([b,a],[b,a])}$ the action profile $([a,b],[a,b])$ is a Nash
  equilibrium and leads to the outcome $a$, however $H([a,b],[a,b]) =
  b$. Hence, $g^H$ does not $NE$-implement $H$.
  
  %% It is not difficult to see that for the function $H$, no truthful
  %% $NE$-implementation can be also an $NE$-implementation, since if a
  %% game form $h$ truthfully $NE$-implements $H$, all the outcomes of
  %% $h$ are determined (the bold outcomes must agree with $H$), but
  %% since this in turn determines the Nash equilibria in any game
  %% $\tuple{h^H,<}$, the problematic case $\textdied\ $ remains a
  %% counter example against $h$ being an $NE$-implementation.
  
  Let us next consider the social choice function $J$ which is
  dictatorial for player $1$, i.e., $J$ is defined by
  
  \begin{itemize*}
  \item[] $J([a,b],[a,b]) = J([a,b],[b,a]) = a$;
  \item[] $J([b,a],[a,b]) = J([b,a],[b,a]) = b$.
  \end{itemize*}
  
  The four possible games $\tuple{g^J,<}$ for $J$ are depicted in
  Figure~\ref{fig:ne-exampltwo}. It is easy to see that the circled
  outcomes in those games are Nash equilibria: they give the preferred
  outcome for $1$ (so $1$ cannot improve by deviating) and they are
  the same in a fixed row (so $2$ cannot change the
  outcome). Moreover, it is also a straightforward check that for all
  those Nash equilibria, the outcome in the game $\tuple{g^J,<}$ is
  the same as $J(<)$ (for instance, in the top left game, both
  equilibria yield $a$ which coincides with $J([a,b],[a,b])$, and in
  the bottom left game, both equilibria yield $b = J([b,a],[a,b])$,
  etc): this justifies the ticks \checked. So $g$ $NE$-implements
  $J$. To show that $g$ also truthfully $NE$-implements $J$, we need
  to check that all the bold outcomes in Figure~\ref{fig:ne-exampltwo}
  are circled and ticked \checked.
  
   \begin{figure}%[h]
   \begin{center}
     \framebox[\linewidth] {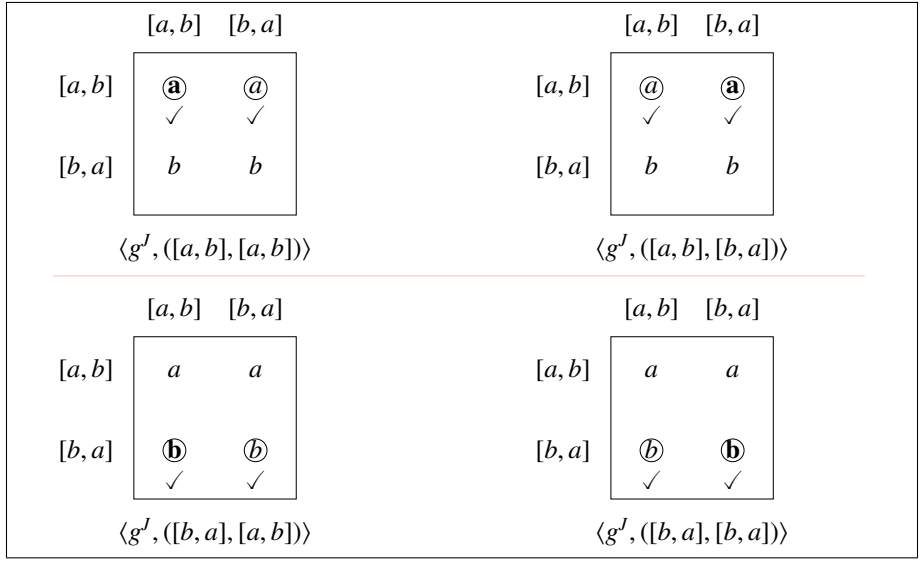 }
   \end{center}
   \caption{ 
      $g^J$ both $NE$-implements and truthfully $NE$-implements $J$.}
      \label{fig:ne-exampltwo}
   \end{figure}

Next, to give an example of an $NE$-implementation that is not a
truthful one, consider the game form $g^J_-$. It is mathematically
equivalent to the game form $g^J$: the outcomes $a$ and $b$ are only
inverted. Playing $g^J_-$, the player $1$ would simply play the
contrary to his true preference. This always yields a Nash equilibrium
and the outcomes are always as prescribed by $J$. Hence, like $g^J$,
the game form $g^J_-$ is an $NE$-implementation of the social choice
function $J$. However, since the player $1$ needs to trick the game in
order achieve a Nash equilibrium, it is easy to see that $g^J_-$ does
not truthfully $NE$-implement $J$. The crosses \textdied\ on
Figure~\ref{fig:ne-examplthree} mark the action profiles that
correspond to the true preferences of the players, and we can see that
their respective outcome always fails to be as prescribed by $J$.

   \begin{figure}%[h]
   \begin{center}
     \framebox[\linewidth] {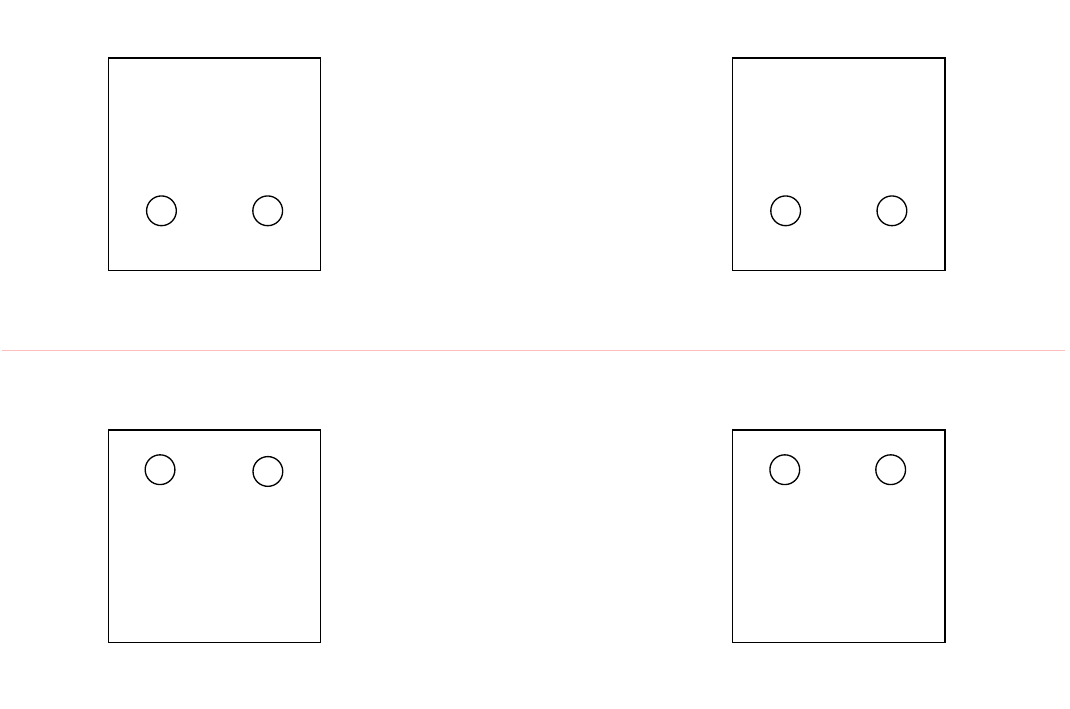 }
   \end{center}
   \caption{ $g^J_-$ $NE$-implements $J$ but does not truthfully
     $NE$-implement it.}
      \label{fig:ne-examplthree}
   \end{figure}

Finally, we argue that it is possible for a game form to be neither a
$NE$-implementation nor a truthful implementation of a given function:
take $P$ such that $P(<) = a$ for all profiles $<$. Moreover, for all
$<$, let all outcomes in the matrix for $\tuple{g^P,<}$ be $b$. For
every $<$, every outcome in $\tuple{g^M,<}$ is a Nash equilibrium (no
agent can change the outcome, let alone improve it). At the same time,
for all $a^*$ we have $b = o(a^*) \neq P(<) = a$, which shows that $g$
does not $NE$-implement $P$. It does also not truthfully
$NE$-implement it: take, for any $<$, a profile $a^*$ in the game
$\tuple{g^M,<}$ such that $a^*\ =\ <$. We have already seen that
$o(a^*) \neq P(<)$, which proves our claim.
\end{example}

%%%%%%%%%%%%%%%%%%%%%%%%%%%%%%%%%%%%%
%%%% LOGIC OF SOCIAL CHOICE FUNCTIONS
\section{A Logic of social choice functions}
\label{sec:lscf}

Following the tradition in implementation theory (cf.\ Remark
\ref{rem:moulin}), we model social choice functions as a particular
kind of strategic game form. In \cite{TrvdHWo09aamas} we proposed a
logic for modelling strategic games on the basis of \CLPC. Every
player controls a set of propositional variables and a strategy for a
player amounts to choosing a truth value for the variables he
controls.  We adapt the ideas of \cite{TrvdHWo09aamas} to game forms
where the strategies of the players correspond to the reports of
preferences.

%%\medskip\noindent\textbf{Semantics.}~
\subsection{Semantics}

Let $X$ be an arbitrary set of propositions. We can see a
\emph{valuation} of $X$ as a subset $V \subseteq X$ where ${\tt tt}$
(i.e., true) is assigned to the propositions in $V$ and ${\tt ff}$
(false) is assigned to the propositions in $X \setminus V$. We denote
the set of possible valuations over $X$ by $\Theta^X$.

In the presence of a set of players $N$ and a set of outcomes $K$, the
set of propositions controlled by a player $i \in N$ is defined as
$At[i,K] = \{ \considers{i}{x}{y} \mid x, y \in K \}$. Every
$\considers{i}{x}{y}$ is a proposition controlled by the agent $i$
which means that $i$ reports that it values the outcome $x$ at least
as good as $y$. We also define $At[N,K] = \cup_{i \in N} At[i,K]$,
which is then the set of all controlled propositions.

We can `encode' a particular preference (or linear order) of player
$i$ as a valuation of the propositions in $At[i,K]$. However,
conversely, not all valuations correspond to a linear order
preference. A strategy of a player $i$ consists of reporting a
valuation of $At[i,K]$ encoding a linear order over $K$. For every
player $i$, we define $strategies[i,K]$ as a set of valuations $V \in
\Theta^{At[i,K]}$ such that: (\emph{i}) $\considers{i}{x}{x} \in V$,
(\emph{ii}) if $x \neq y$ then $\considers{i}{x}{y} \in V$ iff
$\considers{i}{y}{x} \not\in V$, and (\emph{iii}) if
$\considers{i}{x}{y} \in V$ and $\considers{i}{y}{z} \in V$ then
$\considers{i}{x}{z} \in V$.
\begin{remark}
Every $\considers{i}{x}{y}$ could be seen as a predicative expression
$p(i,x,y)$ that would read that agent $i$ reported to prefer the
outcome $x$ over $y$. However, since $N$ and $K$ are finite, we look
at these expressions as a finite collection of propositions. The
constraints of control in Figure \ref{fig:ax-scf} will be their
propositional theory corresponding to the three preceding constraints
on the valuations.
\end{remark}

For every coalition $C \subseteq N$, let $strategies[C,K]$ be the set
of tuples $v_C = (v_i)_{i \in C}$ where $v_i \in strategies[i,K]$. It
is the set of strategies of the coalition $C$. To put it another way,
it corresponds to a valuation of the propositions controlled by the
players in $C$, encoding one preference over $K$ for every player in
$C$.

A \emph{state} (or reported preference profile) is an element of
$strategies[N,K]$, that is, a strategy of the coalition containing all
the players.
We now define the models of social choice functions.
\begin{definition}[Model of social choice functions]\label{def:model-scf}
  A \emph{model of social choice functions over $N$ and $K$} is a
  tuple $M = \tuple{N, K, out, (<_i)}$, such that:
  \begin{itemize*}
  %\item $N = \{1, \ldots , n \}$ is a finite nonempty set of players;
  %\item $K$ is a finite nonempty set of outcomes;
  \item[] $out: strategies[N,K] \longrightarrow K$ maps every state to an
    outcome;
  \item[] For every $i \in N$, $<_i\ \in L(K)$ is the true order of
    preferences of $i$.
  \end{itemize*}
\end{definition}

Hence, every player $i$ has two levels of preferences: (\emph{i}) a
true one, given by $(<_i)$ and (\emph{2}) a reported one, given by a
valuation in $strategies[i,K]$.

Taking out the true preference profile from a model of SCF, we obtain
a mere instantiation of a pre-Boolean game \cite{bonzon07knaw}. It is
required to assign every variable to one (actual control) and only one
(exclusive control) player, but there are some constraints on the
possible valuations (`non-full' control). In \cite{bonzon07knaw},
actual and exclusive control are grasped by an assignment function
(mapping every propositional variable to exactly one player), and the
partial control is modelled by a \emph{set of constraints} given
as a set of satisfiable propositional formulae.

The language $\mathcal{L}^{scf}[N,K]$ is inductively defined by the
following grammar:
$$ 
\begin{array}{lcccccccccccccl}
   \varphi & \assign & \top & \vsep & p & \vsep & x & \vsep & \lnot\varphi & \vsep & \varphi \lor \varphi & \vsep & \Diamond_C \varphi & \vsep & \pospref{i}\varphi\\
 \end{array}
$$ where $p$ is atom of $At[N,K]$, $x$ is an atom of $K$, $i \in N$,
and $C$ is a coalition. Given a model $M$ and a state (i.e., a reported profile $v$), formula $\Diamond_C\varphi$ reads that
provided that the players outside $C$ hold on to their current
strategy $v_C$, the coalition $C$ has a strategy, i.e., a way to announce their profiles, such that $\varphi$ holds. Formula
$\pospref{i}\varphi$ reads that $i$ locally (at the current reported
profile) considers a reported profile where $\varphi$ is true at least
as preferable.

\begin{definition}[Truth values of \mbox{${\mathcal
      L}^{scf}[N,K]$}]\label{def:truth-in-scf} Given a model $M =
  \tuple{N, K, out,%\linebreak 
(<_i)}$, we are going to interpret
  formulae of ${\mathcal L}^{scf}[N,K]$ in a state of the model. A
  state $v = (v_1, \ldots , v_n)$ in $M$ is a tuple of valuations $v_i
  \in strategies[i,K]$, one for each agent. The truth definition is
  inductively given by:
  \begin{center}
    \begin{tabular}{lcl}
%      $M, v \models \top$ & & \\
      $M, v \models p$ & iff & $p \in v_i$ for some $i \in N$\\
      $M, v \models x$ & iff & $out(v) = x$\\
      $M, v \models \lnot \varphi$ & iff & $M, v \not\models
      \varphi$\\
      $M, v \models \varphi \lor \psi$ & iff & $M, v \models
      \varphi$ or $M, v \models \psi$\\
      $M, v \models \Diamond_C\varphi$ & iff & there is a state $u$ such that\\
&& $v_i = u_i$ for every $i \not\in C$ and $M, u \models \varphi$\\
      $M, v \models \pospref{i}\varphi$ & iff & there is a state $u$ such that\\
&& $out(v) <_i out(u)$ and $M, u \models \varphi$\\
      % \multirow{2}{*}{$M, v \models \Diamond_C\varphi$} &
      % \multirow{2}{*}{iff} & there is a state $u$ such that\\
      % & & $v_i = u_i$ for every $i \not\in C$ and $M, u \models \varphi$\\
      % \multirow{2}{*}{$M, v \models \pospref{i}\varphi$} &
      % \multirow{2}{*}{iff} & there is a state $u$ such that\\
      % & & $out(v) <_i out(u)$ and $M, u \models \varphi$
    \end{tabular}
  \end{center}
\end{definition}

We assume that player $i$ only makes claims or announcements about its own preferences, and $i$ controls nothing else, so the atomic clause could equivalently have read
\[
M, v \models p^i_{x > y} \mbox{ iff } p^i_{x > y} \in v_i
\]

The truth of $\varphi$ in all models over a set of players $N$ and a
set of outcomes $K$ is denoted by $\models_{\Lambda^{scf}[N,K]}
\varphi$.  The classical operators $\land$, $\imp$, $\iff$ can be
defined as usual. We also define $\Box_C\varphi \triangleq \lnot
\Diamond_C \lnot \varphi$ and $\necpref{i}\varphi \triangleq \lnot
\pospref{i} \lnot\varphi$.

\begin{theorem}[Decidability]
  The problem of deciding whether a formula $\varphi \in
  \mathcal{L}^{scf}[N,K]$ is satisfiable is decidable.
\end{theorem}
\smallskip
\begin{pf}
  It suffices to remark that $N$ and $K$ are finite. Hence, we can
  enumerate every model of SCF over $N$ and $K$ and check whether
  $\varphi$ is satisfiable in one state of one model.

%  $\hfill\blacksquare$
\end{pf}

%%\medskip\noindent\textbf{Ballots.}~
\subsection{Ballots}
\label{sec:ballot}

We think of a particular preference of $L(K)$ encoded in the
language of the propositions as a \emph{ballot}.

\begin{definition}[Ballot]\label{def:ballot}
  For every player $i \in N$, we can see every $<_i\ \in L(K)$ as a
  permutation $[x_1, x_2 \ldots]$ of the elements of $K$, where the
  more to the left the outcome is, the more it is preferred by the
  player $i$. We can reify in the language the reported preferences,
  that is, the ballot casted by the player $i$: $$\ballot_i(<)
  \triangleq \considers{i}{x_1}{x_2} \land \considers{i}{x_2}{x_3}
  \land \ldots \considers{i}{x_{|K| -1}}{x_{|K|}}.$$ Then, the
  formula $$\ballot(<) \triangleq \bigwedge_{i\in N} \ballot_i(<)$$ is
  a reification of the reported preference profile $<\ = {(<_1, \ldots
    , <_n)}$, consisting of one ballot for every player $i \in N$.
\end{definition}

\begin{remark}\label{rem:hyb-log}
Note that for every $<\ \in L(K)$, the formula $\ballot(<)$ is true at
one and only one state. The reader familiar with Hybrid Logic
\cite{arecescate06hybridlogic} may think of the formula $\ballot(<)$
as a \emph{nominal}, viz.\ a state label available in the object
language.
\end{remark}

\begin{example}\label{ex:ballot}
  Suppose that $N = \{ 1,2 \}$ and $K = \{a,b,c\}$. Let a preference
  profile\linebreak ${(<_1^{ex}, <_2^{ex})} \in L(K)^N$ given by the data of the
  two permutations $[a,c,b]$ and $[c,a,b]$ representing respectively
  the preferences of player $1$ and $2$. This reported preference
  profile can be represented in the language $\mathcal{L}^{scf}[\{ 1,2
  \}, \{a,b,c\}]$ by the formula $$\ballot(<^{ex}) \triangleq
  \considers{1}{a}{c} \land \considers{1}{c}{b} \land
  \considers{2}{c}{a} \land \considers{2}{a}{b}.$$ It is easy to
  verify that the constraints on the elements of $strategies[1,K]$ and
  \linebreak $strategies[2,K]$ are sufficient for inferring a complete
  characterisation of the preference profile. The following is valid
  in the models of social choice functions over $\{ 1,2 \}$ and
  $\{a,b,c\}$:

\begin{center}\begin{tabular}{lcp{9cm}}
\hspace{-0.4cm}    $\ballot(<^{ex})$ & $\iff$ & $\considers{1}{a}{a} \land \considers{1}{b}{b} \land \considers{1}{c}{c} \land \considers{1}{a}{c} \land \considers{1}{c}{b} \land \considers{1}{a}{b} \land  \lnot   \considers{1}{c}{a} \land \lnot \considers{1}{b}{c} \land \lnot \considers{1}{b}{a} \land \considers{2}{a}{a} \land \considers{2}{b}{b} \land \considers{2}{c}{c} \land \considers{2}{c}{a} \land \considers{2}{a}{b} \land \considers{2}{c}{b} \land \lnot \considers{2}{a}{c} \land \lnot \considers{2}{b}{a} \land \lnot \considers{2}{b}{c}$
\end{tabular}\end{center}
\end{example}

%%\medskip\noindent\textbf{Characterising an SCF.}~
\subsection{Characterising an SCF}

Recall that a model of social choice functions is a tuple $M = \tuple{N, K, out, (<_i)}$, where $<_i$ are the real preferences of the agents and the outcome function $o$ assigns to every valuation an element of $K$. There is a one-one correspondence between valuations and preference profiles: the preference profile $P(v)$ associated with valuation $v$ is the relation $<$ for which $x >_i y$ iff $p^i_{x>y} \in v$. Likewise, the valuation $V(<)$ associated with $<$ is the set $\{p^i_{x>y} \mid x >_i y\}$, which collect all the atoms form $\ballot(<)$. This makes it possible to relate a model $M$ with a social choice function $F$ as follows. 

We say that a model $M = \tuple{N, K, out, (<_i)}$ and social choice
function $F: L(K)^N \rightarrow K$ correspond, if for every strategy
profile $<$ and its associated valuation $v$ (i.e., for which $V(<) =
v$ and $P(v) = <$), we have $o(v) = F(<)$.

This correspondence can be syntactically defined in a formula
$\rho^F$:

\[
\rho^F = \bigwedge_{< \in L(K)^N} \Diamond_N(\ballot(<) \land F(<))
\]

Note
that $\Diamond_N$ plays the role of the universal/global existential
modality often noted $E$ in the literature in modal logic:
it allows us to quantify over all the possible valuations in
$\Theta^{At[N,K]}$, or ballots.

Given the outcomes $K$, the agents $N$ and the social choice function $F$, formula $\rho^F$ says that every profile $<$ together with $F(<)$ as an outcome appears in the model. Since the states of a model are all possible profiles in $L(K)^N$, and every profile occurs exactly once, we might as well have defined $\rho^F$ as

\[
\rho^F = \bigwedge_{< \in L(K)^N}(\ballot(<) \rightarrow F(<))
\]

%\begin{definition}[SCF characterisation]
 % We say that $\rho^F \in $\ ${\mathcal L}^{scf}[N,K]$
 % \emph{characterises} the social choice function $F$ if for all $<\ 
 % \in L(K)^N$ and $x \in K$:
%
 % $$F(<) = x \mbox{ iff } \models_{\Lambda^{scf}[N,K]} \rho^F \imp
 % \Diamond_N (\ballot(<) \land x).$$
%\end{definition}

It is easy to see that the logic is expressively complete wrt.\ social
choice functions. That is, for every SCF $F$ over a set of players $N$
and a set of outcomes $K$, there exists a formula $\rho^F \in
{\mathcal L}^{scf}[N,K]$ characterising it. Even though it may not be
optimal in terms of succinctness, it suffices to consider the
conjuncts of formulae $\Diamond_N (\ballot(<) \land x)$, for $<\ \in
L(K)$ and $F(<) = x$. The next example shows, using a simple scenario,
that we can sometimes obtain less na{\"i}ve and more compact characterisations.

\begin{example}\label{ex:n=3:k=2}
  Consider the following model of SCF (or game form) where player $1$
  chooses rows, player $2$ chooses columns and player $3$ chooses
  matrices. There are two outcomes $a$ and $b$. Hence, every
  player $i$ controls the set of atoms $\{ \considers{i}{a}{a},
  \considers{i}{b}{b}, \considers{i}{a}{b}, \considers{i}{b}{a}
  \}$. Every player $i$ has two strategies: $\considers{i}{a}{a} \land
  \considers{i}{b}{b} \land \considers{i}{a}{b} \land \lnot
  \considers{i}{b}{a}$ and $\considers{i}{a}{a} \land
  \considers{i}{b}{b} \land \lnot \considers{i}{a}{b} \land
  \considers{i}{b}{a}$, that we denote respectively by $[a,b]$ and
  $[b,a]$. (In the logic $\Lambda^{scf}[\{ 1, 2,3\}, \{ a,b\}]$, they
  are in fact equivalent to the formulae $\considers{i}{a}{b}$ and
  $\considers{i}{b}{a}$, respectively.)
  \begin{center}
    \begin{tabular}{clcl}
      $[a,b]$ &
      \framebox[0.35\linewidth]{
        \hspace{0.5cm}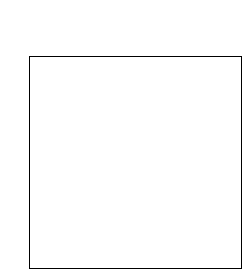
      }
        % {\small
        %   \begin{tabular}{l|c|c}
        %     & $[a,b]$ & $[b,a]$\\
        %     \hline
        %     $[a,b]$ & $a$ & $a$\\
        %     \hline $[b,a]$ & $a$ & $b$
        %   \end{tabular} } }

      &

      $[b,a]$  &
      \framebox[0.35\linewidth]{ 
        \hspace{0.5cm}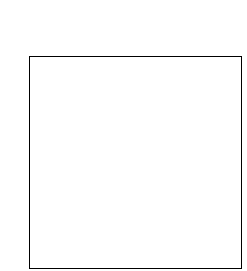
      }
        % {\small
          %   \begin{tabular}{l|c|c}
          %     & $[a,b]$ & $[b,a]$\\
          %     \hline
          %     $[a,b]$ & $a$ & $b$\\
          %     \hline $[b,a]$ & $b$ & $b$
          %   \end{tabular} }

    \end{tabular}
  \end{center}

  We can represent it in the logic $\Lambda^{scf}[\{ 1, 2,3\}, \{
  a,b\}]$ of social choice functions by the formula:
  $$\rho^F \triangleq a \iff (\considers{1}{a}{b} \land \considers{2}{a}{b}) \lor 
  (\considers{1}{a}{b} \land \considers{3}{a}{b}) \lor
  (\considers{2}{a}{b} \land \considers{3}{a}{b}).$$ Note that since
  $out$ is functional, in the models of social choice functions with
  $K = \{ a, b\}$ the outcome $b$ will hold whenever $a$ does not.
  
  Going back to the social choice functions of Example~\ref{ex:first}, we invite the reader to check that
  
  \begin{itemize}
  \item[] $\rho^H = b \leftrightarrow (p^1_{b>a} \land p^2_{b>a})$
  \item[] $\rho^J = a \leftrightarrow p^1_{a>b}$
%  \item[] $\rho^M = a \leftrightarrow (p^1_{a>b} \leftrightarrow p^2_{a>b})$
  \item[] $\rho^P = a$
  \end{itemize} 
\end{example}

%%\medskip\noindent\textbf{True preferences.}~
\subsection{True preferences}

In Section \ref{sec:ballot} we saw how to use the atoms in $At[i,K]$
to encode the reported preference or ballot of a player $i$. These
atoms do not necessarily represent the true preferences of the
agents. We handle the true preferences of player $i$ via the
$\pospref{i}$ modality.

From our basic language $\mathcal{L}^{scf}[N,K]$, we can also define
an operator of interest concerning preferences. We can define the
\emph{global binary} operator of preferences $\psi
\blacktriangleleft_i \varphi$, corresponding to a preference between
propositions. It reads ``all $\varphi$ are better than all $\psi$''.
$$\psi \blacktriangleleft_i \varphi \triangleq \Box_N \bigvee_{<\ \in
  L(K)^N} (\ballot(<) \land (\varphi \imp \Box_N ( \psi \imp
\pospref{i} \ballot(<) ).$$ 
Agent $i$ judges the proposition
$\varphi$ at least as good as $\psi$ iff when the reported preference profile is $<$
and $\varphi$ holds at the state labeled by $\ballot(<)$, then,
whenever $\psi$ holds in a state, $i$ would prefer the state labeled
by $\ballot(<)$ (cf.\ Remark \ref{rem:hyb-log}).

%%\medskip

As in Definition \ref{def:ballot} for reported preferences, we can now
reify the true preferences. Provided that $x$ and $y$ are two possible
outcomes, the formula $y \blacktriangleleft_i x$ captures the fact the
player $i$ prefers (globally) the outcome $y$ over the outcome
$x$. Hence, from a preference profile $<\ \in L(K)^N$, we reify the
preference $[x_1, x_2 \ldots]$ of the player $i$ as follows:
%$$\trueprofile_i(<) \triangleq 
%\Diamond_N (x_2 \land \pospref{i}x_1)\land 
%\Diamond_N (x_3 \land \pospref{i}x_2)\land \ldots 
%\Diamond_N (x_{|K|} \land \pospref{i}x_{|K| -1}).$$ 

$$\trueprofile_i(<) \triangleq
(x_{|K|} \blacktriangleleft_i x_{|K| -1}) \land \ldots \land (x_3 \blacktriangleleft_i x_2) \land
(x_2 \blacktriangleleft_i x_1).$$
% It is equivalent to $\lnot (x_1 <_i x_2) \land \lnot (x_2 <_i x_3)
% \land \ldots \lnot (x_{|K| -1} <_i x_{|K|})$.

Then, the formula $$\trueprofile(<) \triangleq \bigwedge_{i\in N}
\trueprofile_i(<)$$ is a reification of the true preference profile $<
= {(<_1, \ldots , <_n)}$.

\begin{remark}\label{rem:lang-weak}
  Whenever in a model of social choice function $M$ the true
  preference of a player $i$ is such that $x <_i y$, then the formula
  $x \blacktriangleleft_i y$ is true at every state of $M$. However,
  the other way around does not hold. Indeed, when either $x$ or $y$
  is not a possible outcome of a model, the formula $x
  \blacktriangleleft_i y$ is always true for every $i$. From the
  definition,
$x \blacktriangleleft_i y \triangleq \Box_N \bigvee_{<\ \in L(K)^N}
  ({\ballot(<)} \land (y \imp \Box_N ( x \imp \pospref{i} {\ballot(<)}
  ).$ Hence, if $y$ is not a possible outcome, the main implication
  $y \imp \Box_N ( x \imp \pospref{i} \ballot(<) )$ is always true for
  $y$ being always false. Likewise, if $x$ is not a possible outcome, the
  implication $x \imp \pospref{i} \ballot(<)$ is always true for $x$
  being always false. In turn, it makes the main implication always
  true. Also, $\bigvee_{<\ \in L(K)^N} \ballot(<)$ will always be
  satisfied since a state of evaluation represents a ballot by
  definition.

  The object language does not allow to talk about true preferences on
  impossible outcomes. This observation will have a consequence in the
  way we prove the completeness of the logic.
\end{remark}

%%\medskip\noindent\textbf{Axiomatics.}~
\subsection{Axiomatics}

The axiomatization of the models of social choice functions is
presented in Figure \ref{fig:ax-scf}.

\begin{figure*}[ht]
\small
  \framebox[\linewidth]{\hspace{-2cm}
    \begin{tabular*}{0.8\linewidth}{p{.20\linewidth}l}
      \multicolumn{2}{l}{{\bf Constraints of control}}  \\
      ($refl$) & $\considers{i}{x}{x}$\\
      ($\mbox{\it antisym-total}$) & $\considers{i}{x}{y} \iff \lnot \considers{i}{y}{x}$ \hfill {\small , where $x \neq y$}\\
      ($trans$) & $\considers{i}{x}{y} \land \considers{i}{y}{z} \imp \considers{i}{x}{z}$\\
      \multicolumn{2}{l}{{\bf Propositional control}}\\
      ($Prop$) & $\varphi$ \hfill {\small , where $\varphi$ is
        a propositional tautology}\\ 
      ($K(i)$) & $\Box_i (\varphi \imp
      \psi) \imp (\Box_i \varphi \imp \Box_i \psi)$\\ 
      ($T(i)$) & $\Box_i\varphi \imp \varphi$\\ 
      ($B(i)$) & $\varphi \imp \Box_i\Diamond_i \varphi$\\ 
      ($comp\cup$) & $\Box_{C_1}\Box_{C_2}\varphi \iff 
      \Box_{C_1 \cup C_2}\varphi$\\ 
      ($confl$) & $\Diamond_i \Box_j \varphi \imp \Box_j \Diamond_i \varphi$\\
      ($empty$) & $\Box_\emptyset \varphi \iff \varphi$\\ 
      % EXCLU
      ($exclu$) & $(\Diamond_i p \land \Diamond_i \lnot p) \imp 
      (\Box_j p \lor \Box_j \lnot p)$ \hfill {\small , where $j \not = i$}\\
      % FULL
      ($ballot$) & $\Diamond_i \ballot_i(<)$\\
      ($\mbox{\it comp-At}$) & $\Diamond_{C_1} \delta_1 \land \Diamond_{C_2} \delta_2 \imp \Diamond_{C_1 \cup C_2} (\delta_1 \land \delta_2)$\\
      \multicolumn{2}{l}{{\bf Outcomes and preferences}}  \\
      ($func1$) & $\bigvee_{x \in K} (x \land \bigwedge_{y \in K
        \setminus \{ x \}} \lnot y)$\\ 
      ($func2$) & $(\ballot(<) \land \varphi) \imp \Box_N(\ballot(<) \imp \varphi)$\\
      ($incl$) & $\Box_N\varphi \imp \necpref{i}\varphi$\\
      ($K(<_i)$) & $\necpref{i} (\varphi \imp \psi)
      \imp (\necpref{i} \varphi \imp \necpref{i}
      \psi)$\\ 
%      ($T(<_i)$) & $\varphi \imp \pospref{i}\varphi$\\
      ($4(<_i)$) & $\pospref{i}\pospref{i}\varphi
      \imp \pospref{i}\varphi$\\
%      ($refl'$) & $x <_i x$\\
%      ($antisym'$) & $(x <_i y) \iff \lnot (y <_i x)$ \hfill {\small , where $x \neq y$}\\
      ($antisym'$) & $(\ballot(<) \land \pospref{i} \ballot(<') \imp 
      \Box_N (\ballot(<') \imp \necpref{i} \lnot \ballot(<)$\\
%      ($trans'$) & $(x <_i y) \land (y <_i z) \imp (x <_i z)$\\
%      ($total'$) & $\lnot (x <_i y) \imp (y <_i x)$\\
      ($total'$) &  $(\ballot(<) \land \pospref{i} \ballot(<') \lor 
      \Box_N (\ballot(<') \imp \pospref{i} \ballot(<)$\\
      ($unifPref$) & $(x \land \pospref{i}y) \imp (x \blacktriangleleft_i y)$\\
      {\bf Rules} & \\
      ($MP$) & from $\vdash \varphi \imp \psi$ and 
      $\vdash \varphi$ infer $\vdash \psi$\\
      ($Nec(\Box_i)$) &  from $\vdash \varphi$ infer $\vdash \Box_i \varphi$\\
    \end{tabular*}
  }
  
  \caption{{\em Logic of social choice functions
      $\Lambda^{scf}[N,K]$.} $i$ ranges over $N$, $C_1$ and $C_2$ over
    $2^N$, $x$ and $y$ are over $K$, and $<$ is over
    $L(K)^N$. $\delta_1$ and $\delta_2$ are two formulae from
    $\mathcal{L}^{scf}[N,K]$ that do not contain a common atom from
    $At[N,K]$. $\varphi$ represents an arbitrary formula of
    $\mathcal{L}^{scf}[N,K]$, and $p$ an arbitrary atom in
    $At[N,K]$.}\label{fig:ax-scf}
\end{figure*}

Constraints of control ($refl$), ($\mbox{\it antisym-total}$) and
($trans$) say that every player casts an appropriate valuation of its
controlled atoms: a valuation must encode a linear order. ($comp\cup$)
defines the local ability of coalitions in terms of local abilities of
sub-coalitions. The transitivity of the operator $\Box_C$ is the
consequence of ($comp\cup$). Hence, together with ($T(i)$) and
($B(i)$), it makes of $\Box_C$ an $S5$ modality. ($empty$) means that
the empty coalition has no power. ($comp\cup$) and ($confl$) together
make sure that the agents' choices are independent. ($exclu$) means
that if an atom is controlled by a player $i$, the other players
cannot change its value. ($ballot$) makes sure that an agent is always
locally able to cast any preference. From ($\mbox{\it comp-At}$),
provided that $\delta_1$ and $\delta_2$ do not contain a commonly
controlled atom, if a coalition $C_1$ can locally enforce $\delta_1$
and $C_2$ can locally enforce $\delta_2$ then they can enforce
$\delta_1 \land \delta_2$ together.

Axiom ($func1$) forces the fact that for every action profile there is
one and only one outcome. ($func2$) ensures that the outcomes are only
determined by the valuations. ($incl$) ensures that if something is
settled, a player cannot prefer its negation. ($4(\prefeq_i)$)
characterises transitivity. ($antisym'$) and ($total'$) force that the
relation of preference over states is antisymmetric and total (and
hence, in particular, this relation is reflexive).  Finally,
($unifPref$) specifies a fundamental interaction between preferences
and the outcomes. If the casted preference profile at hand leads to
$x$ and agent $i$ prefers an action profile leading to $y$, then at
every action profile leading to $x$, agent $i$ will prefer every
action profile leading to $y$, that is, all $y$ are better than all
$x$.

The logic has a clear flavour of normal modal logic
\cite{ChellasML}. The presence of ($K(i)$) with the necessitation
rule ($Nec(\Box_i)$) gives to the operator $\Box_i$ the property of
normality. The necessitation rule for the operator $\necpref{i}$ holds
because of ($Nec(\Box_i)$) and the axioms ($comp\cup$) and
($incl$). The normality of the modality $\necpref{i}$ then follows
from ($K(<_i)$).

%%\medskip

The axiomatics is largely inspired by the axiomatics of the logic of
games and propositional control (henceforth LGPC) presented in
\cite{TrvdHWo09aamas}. The logic LGPC is designed to model strategic
games in general. The agents have arbitrary strategies, and
preferences allowing for indifference between two different
outcomes. On the other hand, in this paper we focus on SCFs and
hence on particular strategic games that `represent' an SCF (cf.\
Remark \ref{rem:moulin}).

While in LGPC we had an axiom saying that every atom was actually
controlled by at least one agent, here we are more specific as we know
\emph{a priori} which atoms are controlled by a given agent. This is
the role of the axiom ($ballot$). Constraints of controls are also
specific to the present study. The truth values of the controlled
atoms cannot be independent of each other as we use them to encode
preferences. In LGPC, all valuations of the controlled atoms were
permitted.

\putaway{
\begin{proposition}
  For every $i \in N$, if $\vdash \varphi$ then $\vdash \necpref{i}
  \varphi$.
\end{proposition}

\begin{pf}
  Suppose $\vdash \varphi$. Then $\vdash \Box_1 \varphi$, and $\vdash
  \Box_1 \Box_2 \varphi$ \ldots and $\vdash \Box_1 \Box_2\ldots\Box_n
  \varphi$ by successive use of ($Nec(\Box_i)$). Then, $\vdash \Box_N
  \varphi$ by ($comp\cup$). Finally, by ($incl$) we have $\vdash
  \necpref{i} \varphi$.
\end{pf}
}
% COMPLETENESS THEOREM
\begin{theorem}[Soundness and completeness]\label{th:compl}
  $\Lambda^{scf}[N,K]$ is sound and complete with respect to the class
  of models of social choice functions.
\end{theorem}
\smallskip
\begin{pf}
  The proof of completeness first gives an equivalent but more
  standard semantics to the logic: the Kripke models of SCF. Then we
  build the canonical model. For every consistent formula $\varphi$,
  we show how to isolate a sub-model $M_{\varphi}$ that we prove is a
  Kripke model of SCF that satisfies $\varphi$.

  Further details are given in the Appendix. 

%  $\hfill\blacksquare$
\end{pf}

%%%%%%%%%%%%%%%%
%%% APPLICATIONS

\section{Applications}
\label{sec:applications}

We have already demonstrated that the language allows to completely
characterise an SCF. In this section we show how we can express
properties of social choice functions in the language and apply the
logic to reason about them.

The language can be used to characterise requirements on social choice
functions. We first illustrate that with some simple properties, namely
citizen sovereignty and non-dictatorship. Next, we will characterise a
dominant strategy equilibrium. Finally, we provide a formalisation of
monotonicity and strategy-proofness, and use standard results of SCT
to show how we can use the logic to check whether an SCF is
implementable in a dominant strategy.

%%\medskip\noindent\textbf{Citizen sovereignty and non dictatorship.}~
\subsection{Citizen sovereignty and non dictatorship}

%in Example \ref{ex:n3k2-str}.

We say that an SCF satisfies \emph{citizen sovereignty} iff every
outcome in $K$ is feasible. That is, no outcome is rejected
independently of the individual opinions. It is defined as follows.
\begin{definition}[Citizen sovereignty]
  An SCF $F$ satisfies \emph{citizen sovereignty} iff for every $x \in
  K$ there is a $<\ \in L(K)^N$ such that $F(<) = x$.
\end{definition}
%It says that for every outcome $x$, there is at least a ballot
%whose outcome is $x$.

The next formula is a straightforward translation of the definition of
citizen \linebreak sovereignty in the language of social choice functions.
  $$\CITSOV \triangleq \bigwedge_{x \in K} \Diamond_N x.$$

%\begin{example}%\label{ex:n3k2-str}
%  Consider the model $M$ of SCF in Example \ref{ex:n=3:k=2}. For every
%  outcome $x$, there is a deviation of the coalition $N$ such that
%  $x$ is the case. In other words, the SCF satisfies the property of
%  citizen sovereignty.

%  It is clear that $M, v \models \CITSOV$ for every state $v \in
%  Val[\{1,2,3\}, \{a,b\}]$.
%\end{example}

We say that an SCF satisfies \emph{non dictatorship} iff no player can
always impose its favourite outcome.
\begin{definition}[Non-dictatorship]
  An SCF $F$ is \emph{non dictatorial} iff for every player $i \in N$
  there is a ballot $<\ \in L(K)^N$ such that $F(<) <_i y$ for some $y
  \in K \setminus \{F(<)\}$.
\end{definition}
This says that for every player, there is a ballot $<$ whose outcome is
$F(<)$, and $i$ prefers an outcome that is not $F(<)$.

We can rewrite the definition of non dictatorship into the language of
social choice functions as follows.
  $$\NODICT \triangleq \bigwedge_{i \in N} \Diamond_N \left (\bigvee_{x \in K} 
    \left (x \land \bigvee_{y \in K \setminus \{x\}}
      \considers{i}{y}{x} \right ) \right ).$$

%\begin{example}
%  Again, it is trivial to verify that the same model of SCF in Example
%  \ref{ex:n=3:k=2} satisfies the property of non-dictatorship. It is
%  clear that $M, v \models \NODICT$ for every state $v \in
%  Val[\{1,2,3\}, \{a,b\}]$.
%\end{example}

The following proposition is immediate.
\begin{proposition}
  Consider a social choice function $F$ and $\rho^F$ a formula
  characterising it. 
\begin{enumerate}
\item $F$ has the property of citizen sovereignty iff
  $\models_{\Lambda^{scf}[N,K]} \rho^F \imp \CITSOV$.
\item $F$ is non dictatorial iff $\models_{\Lambda^{scf}[N,K]} \rho^F
  \imp \NODICT$.
\end{enumerate}
\end{proposition}

%%\medskip\noindent\textbf{Dominant strategy equilibrium.}~
\subsection{Dominant strategy equilibrium}

Citizen sovereignty and non dictatorship are possible properties of a social
choice function: their formulations in logic are globally true (or
false) in a model of SCF.
% $\models_{\Lambda^{scf}[N,K]} \CITSOV \imp \Box_N \CITSOV$ and
% $\models_{\Lambda^{scf}[N,K]} \NODICT \imp \Box_N \NODICT$.
However, the logic is also able to formalise solution concepts, which
are properties of states. In \cite{TrvdHWo09aamas}, we 
characterised several solution concepts (dominant strategy
equilibrium, Nash equilibrium, core membership\ldots) that are
directly applicable in the logic of the present work.

In order to formalise strategy-proofness later, we need to
characterise a dominant strategy equilibrium. A dominant strategy
equilibrium captures a particularly important pattern of behaviour. It
arises when every player plays a dominant strategy, that is, a
strategy that would represent the best choice whatever the other
agents play. We define it directly in our models of SCF.

\begin{definition}[Dominant strategy equilibrium]
  Let $v^*$ be a state in a%\linebreak 
model of social choice functions
  ${\tuple{N, K, out, (<_i)}}$. $v^*$ is a \emph{dominant strategy
    equilibrium} iff for every player $i \in N$ and every strategy
 % \linebreak
$u_{N \setminus \{i\}} \in {strategies[N \setminus \{i\},K]}$, we
  have $out(u_0 \ldots u'_i \ldots u_n)\linebreak <_i out(u_0 \ldots v^*_i
  \ldots u_n)$ for every $u'_i \in {strategies[i,K]}$.
\end{definition}

A dominant strategy equilibrium is a strong solution concept: such an
equilibrium does not depend on the knowledge of an agent $i$ about the
strategies or preferences of other players.

It is convenient to introduce the notion of best response by an agent $i$.
$$\BR_i \triangleq \bigvee_{x \in K} (x \land \Box_i\pospref{i}x ).$$
A player $i$ plays a best response in a state if, $x$ being the
outcome, for every deviation of $i$, $i$ prefers $x$.

We can now define strategy dominance in terms of best response:
$$\DOM \triangleq \bigwedge_{i \in N} \Box_{N \setminus \{i\}}\BR_i.$$
We have a strategy dominant state if the current choice of every
player ensures them a best response whatever other agents do.

\begin{proposition}
  Assume a model of social choice functions $M$ and a state $v^*$. We
  have that $v^*$ is a dominant strategy equilibrium iff $M, v^*
  \models \DOM$.
\end{proposition}

%%\medskip\noindent\textbf{Monotonicity and strategy-proofness.}~
\subsection{Monotonicity and strategy-proofness}

%%%%%%%%%%%%%%%%%%%%%%%%%%%%%%%%%%%%%%%%%%%%%%%%%%%%%%%%%%%%%%%%

One important property of SCF is \emph{monotonicity}, as this property
can affect the implementability of social choice functions.
\begin{definition}[Monotonicity]\label{def:monoton}
  An SCF $F$ is \emph{monotonic} iff for all ${\{<,<'\}} \subseteq
  L(K)^N$ and $x \in K$, if $F(<) = x$ and if for all $i \in N$, for
  all $y \in K$ we have that that $y <_i x$ implies that $y <'_i x$,
  then, $F(<') = x$.
\end{definition}

We propose to characterise monotonic social choice functions.
We define
\[
\MON \triangleq 
\left\{
  \begin{array}{c}
    \bigwedge_{<\ \in L(K)^N} \bigwedge_{<'\ \in L(K)^N} 
    \bigwedge_{x \in K}
    \bigl[
    \Diamond_N (\ballot(<) \land x) \land \\
    \bigwedge_{i \in N} \bigwedge_{y \in K} 
    \bigl (
      \Diamond_N (\ballot(<) \land \considers{i}{x}{y}) \imp \\
      \Diamond_N (\ballot(<') \land \considers{i}{x}{y})
    \bigr ) 
    \imp \Diamond_N (\ballot(<') \land x)
    \bigr]
  \end{array}
\right\}.
\]

%%\medskip 

Although it may appear rather complex, the predicate $\MON$ is
essentially nothing more than the expression of Definition
\ref{def:monoton} in our language $\mathcal{L}^{scf}[N,K]$. The
following proposition is immediate.
\begin{proposition}
  Consider a social choice function $F$ and $\rho^F$ a formula
  characterising it. $F$ is monotonic iff
  $$\models_{\Lambda^{scf}[N,K]} \rho^F \imp \MON.$$
\end{proposition}

Monotonicity does not depend on the true preference profile of the
players. Accordingly, our definition does not involve the modalities
of preference $\pospref{i}\varphi$ and $\varphi \blacktriangleleft_i
\psi$. Capitalising on standard results from social choice theory, we
will show that using the full expressivity of our language (that is,
using true preference modalities) we can obtain a much simpler
formulation.

%%\medskip

We say that an SCF is strategy-proof if for every preference profile,
telling the truth (reporting the true preference) is a dominant
strategy for every player.
\begin{definition}[Strategy-proofness]
  An SCF $F$ is \emph{strategy-proof} iff $F$ is truthfully
  $\DOM$-implementable.
\end{definition}
Hence, a choice function is strategy-proof when it is truthfully
implementable in dominant strategy: for every preference profile,
reporting their true preference is a dominant strategy for every
player.

The \emph{revelation principle} \cite{gibbard73} is a central result
in implementation theory. It states that if an SCF is
$\DOM$-implementable, then it is truthfully $\DOM$-implementable. It
is true in general even if $L(K)$ is based on weaker orders.
The revelation principle tells us that if an SCF $F$ is implementable
in dominant strategies then there exists a direct mechanism such that for
every preference profile $<$, truth telling (every player $i$ reports
$<_i$) is a dominant strategy and the outcome is $F(<)$.

Truthful implementations are rather weak; it is easier in general to
implement a choice function truthfully than with `standard'
implementations. Indeed, in truthful implementations there might be an
equilibrium that leads to an outcome different of the one prescribed
by the SCF. But because in this paper we consider linear preferences,
and we assume that players cannot be indifferent between two distinct
outcomes, such a situation cannot happen. Thus, we can be more
specific than the revelation principle.

\begin{theorem}[\mbox{\cite[Corollary 4.1.4]{dasgupta79inc-comp}}]\label{th:equiv}
  A direct mechanism $g$ truthfully implements an SCF $F$ in dominant
  strategies iff $g$ $\DOM$-implements $F$.
\end{theorem}
Hence, when working in dominant strategies with linear preferences, the
concepts of implementation and truthful implementation coincide.

%%\medskip

We propose to characterise strategy-proof social choice functions as follows:
$$\STRPROOF \triangleq 
\bigwedge_{<\ \in L(K)^N} \left [ \trueprofile(<) \imp (\ballot(<) \imp
  \DOM) \right] $$

The formula $\STRPROOF$ is an immediate reformulation of
the definition of strategy-proofness in our language of social choice
functions.
\begin{proposition}
  Consider a social choice function $F$ and $\rho^F$ a formula
  characterising it. $F$ is strategy-proof
 iff
  $$\models_{\Lambda^{scf}[N,K]} \rho^F \imp \STRPROOF.$$
\end{proposition}

This Proposition provides us with a general procedure to check whether
a social choice function is strategy-proof. Moreover (because of
Theorem \ref{th:equiv}), because we restrict our attention to linear
preferences, it allows us to check whether an SCF is
$\DOM$-implementable.

\begin{example}
  We can verify for instance that the social choice function
  characterised in Example \ref{ex:n=3:k=2} is strategy-proof.
  \putaway{
  $$\models_{\Lambda^{scf}[\{ 1, 2,3\}, \{ a,b\}]} (a \iff (\considers{1}{a}{b}
  \land \considers{2}{a}{b}) \lor (\considers{1}{a}{b} \land
  \considers{3}{a}{b}) \lor (\considers{2}{a}{b} \land
  considers{3}{a}{b}) \imp \STRPROOF.$$ }

%\begin{tabular}{p{\linewidth}}
\begin{tabular}{ll}
  $\models_{\Lambda^{scf}[\{ 1, 2,3\}, \{ a,b\}]}$ & $(a \iff {(\considers{1}{a}{b}
    \land \considers{2}{a}{b})} \lor {(\considers{1}{a}{b} \land \considers{3}{a}{b})} \lor {(\considers{2}{a}{b}
    \land \considers{3}{a}{b})})$\\ 
  & $\imp \STRPROOF$.
\end{tabular}

\end{example}

Monotonicity sometimes implies implementability and this is actually
the case in our setting. Since we are working with \emph{rich domains}
of preferences\footnote{The notion of a rich domain is some tangential
  to the purposes of this paper. Briefly, our domain of preferences is
  rich because we allow every linear order of $K$. See \cite[Sec.\
  3.1]{dasgupta79inc-comp}} and linear orderings the following result
holds.

\begin{theorem}[\mbox{\cite[Cor.\ 3.2.3, Th.\ 4.3.1]{dasgupta79inc-comp}}]
\label{th:tru-mon}
An SCF is truthfully implementable in dominant strategies iff it is
monotonic.
\end{theorem}

This standard result of implementation theory shows that in our
setting, the notions of monotonicity and of strategy-proofness
match. Trivially we are actually able to substantially simplify the
formula $\MON$, our characterisation of monotonicity in the formal
language. Indeed, as a consequence of Theorem \ref{th:tru-mon}, we
have the following.
\begin{proposition}\label{prop:ldi-mon-truth}
$$\models_{\Lambda^{scf}[N,K]} \MON \iff \STRPROOF.$$
\end{proposition}

\section{Discussion and perspectives}\label{sec:conclusion}

We have presented the problem of direct implementation in social
choice theory and proposed a logical formalisation of it. We were able
to give a sound and complete axiomatization to the logic.  We showed
how we can characterise social choice functions and properties of
social choice functions. And finally, we have demonstrated the value
of the logic by proposing a general logical procedure for checking
whether a social choice function is strategy-proof.

Our logical language is a formal counterpart of the language of
``natural mathematics'' that is typically used in social choice
theory. There are however two features that make it particularly
useful: (\emph{i}) it is supported by a non ambiguous semantics; and
(\emph{ii}) the resulting logic is decidable.

Section \ref{sec:applications} suggests a logical methodology for
reasoning about problems of social choice theory with the logic of
social choice functions. Let a collection of properties of social
choice theory $Pi, i \in \{1, \ldots n\}$ be characterised in the logic
$\Lambda^{scf}[N,K]$ by $\rho^{Pi}$, respectively.

\begin{enumerate}
\item \emph{We can use the logic in order to check whether an SCF
    satisfies a certain property.} An SCF $F$ characterised by
  $\rho^F$ has the property $P1$ iff $\rho^{F} \imp \rho^{P1}$ is
  derivable in $\Lambda^{scf}[N,K]$.

\item \emph{We can use the logic in order to evaluate the strength of
    constraints in SCT.} $P1$ is a property weaker than $P2$ iff the
  formula $\rho^{P2} \imp \rho^{P1}$ is derivable in
  $\Lambda^{scf}[N,K]$. For instance, instead of using a result of SCT
  to prove Proposition \ref{prop:ldi-mon-truth}, we could actually use
  the logic to automatically verify that monotonicity and
  strategy-proofness coincide in the current setting. More
  interestingly, we could use it to prove new theorems.

\item \emph{We can use the logic for mechanism design.} Building a
  mechanism that implements a social choice procedure satisfying the
  properties $P1, P2, \ldots\linebreak Pn$ consists of finding a model
  for the formula $\rho^{P1} \land \rho^{P2} \land \ldots \land
  \rho^{Pn}$.
\end{enumerate}
We believe these are exciting possibilities for social choice theory
and logic, and as the logic is decidable, they are in principle
possible.

\section*{Acknowledgment}
%\begin{acknowledgements}
An earlier abstract of this paper appeared as \cite{TrvdHWo09tark}.
We thank the anonymous reviewers for their suggestions that helped
to improve the paper. We are also grateful to the
participants of TARK'09. This research is funded by the EPSRC grant
EP/E061397/1 \emph{Logic for Automated Mechanism Design and Analysis
  (LAMDA)}.
%\end{acknowledgements}

\bibliography{lamda-biblio}
\bibliographystyle{plain}

\section*{Proof of Theorem \ref{th:compl}}

% COMPLETENESS THEOREM

$\Lambda^{scf}[N,K]$ is sound and complete with respect to the class
of models of social choice functions.

%\putaway{
  % PROOF
\smallskip
  \begin{pf}
    It is routine to verify that all principles of Figure
    \ref{fig:ax-scf} are valid. We show that if a formula is
    consistent, it is provable in the system $\Lambda^{scf}[N,K]$.

    We first introduce the Kripke models of SCF. A \emph{Kripke model
      of SCF} is a tuple $M = \tuple{N, K, S, (R_i), (P_i),V}$ such
    that:
    \begin{itemize*}
    \item $N$ and $K$ are parameters;
    \item $S = \{ V \in \Theta^{At[N,K]} \mid \forall i \in N, \exists
      V_i \in strategies[i,K] \mbox{ s.t. } V = \cup_{i \in N}  V_i \}$;
    \item $V$ is a valuation function of $At[N,K] \cup K$ where for
      every $v \in S$:
      \begin{itemize*}
      \item $p \in V(v)$ iff $p \in v$, $p \in At[N,K]$;
      \item there is a unique $x \in K$ s.t.\ $x \in V(v)$;
        \emph{[$\hookrightarrow$ we say that the model is \emph{based}
          on the outcome function $out^M$ when $out^M(v) = x$ iff $x
          \in V(v)$]}.
      \end{itemize*}
    \item $R_i vu$ iff $v_j = u_j$ for all $j \neq i$;
    \item there is a $<^M\ \in L(K)^N$ s.t.\ $P_i vu$ iff (if $x \in
      V(v)$ and $y \in V(u)$ then $x <^M_i y$);
      \emph{[$\hookrightarrow$ we say that the model is \emph{based}
        on $<^M$]}.
    \end{itemize*}

    Truth values of $\Diamond_i\varphi$ and $\pospref{i}\varphi$ in a
    Kripke model of SCF are obtained in the standard way from the
    relations $R_i$ and $P_i$, respectively.
  
    Clearly, for every Kripke model $M$ based on $out^M$ and $<^M$, we
    can construct a model of social choice functions $M^{scf} =
    \tuple{N, K,out^M,(<^M_i)}$ and reciprocally.
  
    By construction, there exists a bijection $f: S \longrightarrow
    strategies[N,K]$ that associates a state $s$ in $M$ to a state $v
    = (v_1 \ldots v_n)$ in $M^{scf}$ in such a way that for every $p
    \in At[i,K]$, we have $p \in V(s)$ iff $p \in v_i$.
  
    The following is easy to see.
    \begin{claim}
      $M, s \models \varphi$ iff $M^{scf}, f(s) \models \varphi$.
    \end{claim}
    Hence, the proof of the theorem can be reduced to a proof of
    completeness of the logic wrt.\ to the class of Kripke models of
    SCF.

    %%\medskip

    Let $\Xi$ be the set of maximally consistent sets (mcs.) of
    $\Lambda^{scf}[N,K]$. We define the proper canonical model
    $M^{can} = \tuple{N,K,S,(R_i),(P_i),V}$ as follows. $N$ and $K$
    are the parameters of the logic. $S = \Xi$.
    $R_i \Gamma\Delta$ iff $\forall \delta \in \Delta$, $\Diamond_i
    \delta \in \Gamma$. $P_i\Gamma\Delta$ iff $\forall \delta \in
    \Delta$, $\pospref{i}\delta \in \Gamma$. $p \in V(\Gamma)$ iff $p
    \in \Delta$. $x \in V(\Gamma)$ iff $x \in \Delta$.

    Given an mcs.\ $\Gamma_0$ we define the set of mcs. `describing'
    the same SCF \emph{and} where the players have the same true
    preferences (modulo the preferences concerning some outcome
    which is not feasible in the SCF):

\begin{tabular}{lcp{8cm}}
  $Cluster(\Gamma_0)$ & $\triangleq$ & $\{ \Gamma_1 \mid \forall <\ \in L(K)^N, \forall x \in K, \Diamond_N (\ballot(<) \land x) \in \Gamma_1  \mbox{ iff } \Diamond_N (\ballot(<) \land x) \in \Gamma_0 \} \cap \{ \Gamma_2 \mid \forall i \in N, \forall \{ x, y\} \subseteq K, x \blacktriangleleft_i y \in \Gamma_2  \mbox{ iff } x \blacktriangleleft_i y \in \Gamma_0 \}$
\end{tabular}

\putaway{
    \[
    Cluster(\Gamma_0) =
    \begin{array}{l}
      \{ \Gamma_1 \mid \forall <\ \in L(K)^N, \forall x \in K, \Diamond_N (\ballot(<) \land x) \in \Gamma_1  \mbox{ iff } \Diamond_N (\ballot(<) \land x) \in \Gamma_0 \}\\
      \cap \{ \Gamma_2 \mid \forall i \in N, \forall \{ x, y\} \subseteq K, x \blacktriangleleft_i y \in \Gamma_2  \mbox{ iff } x \blacktriangleleft_i y \in \Gamma_0 \}$$\\
    \end{array}.
    \]
}
%\medskip

    Let $\varphi$ be a consistent formula of
    $\mathcal{L}^{scf}[N,K]$. There is an mcs.\ $\Gamma_\varphi$ s.t.\
    $\varphi \in \Gamma_\varphi$. The proof consists in constructing a
    model from $\Gamma_\varphi$ such that it is indeed a Kripke model
    of SCF and there is a state satisfying $\varphi$.

    We define $M_\varphi = \tuple{N',K',S',R'_i,P'_i,V'}$ from
    $M^{can}$ as follows:
    \begin{itemize*}
    \item $N' = N$ and $K' = K$;
    \item $S' = \Xi_{\mid Cluster(\Gamma_\varphi)}$;
    \item $R'_i = {R_i}_{\mid Cluster(\Gamma_\varphi)}$;
    \item $P'_i = {P_i}_{\mid Cluster(\Gamma_\varphi)}$;
    \item $p \in V'(\Delta)$ iff $p \in V(\Delta)$, $\Delta \in S'$.
    \end{itemize*}

    It is immediate that the truth lemma holds.
    \begin{claim}
      $M_\varphi, \Gamma \models \delta$ iff $\delta \in \Gamma$.
    \end{claim}    
    Hence, $M_\varphi, \Gamma_\varphi \models \varphi$.

    The set of states in Kripke models of SCF is defined as the set of
    valuations of $At[N,K]$ encoding a preference profile. We prove
    that there exists a bijection between $S'$ and $L(K)^N$.
    \begin{claim} The following statements are true:
      \begin{enumerate}
      \item $\forall \Delta \in S', \exists ! <\ \in L(K)^N$ s.t.\
        $\ballot(<) \in \Delta$;
      \item $\forall <\ \in L(K)^N, \exists ! \Delta \in S'$ s.t.\
        $\ballot(<) \in \Delta$.
      \end{enumerate}
    \end{claim}
    The first part of the claim follows from the constraints of control
    ($refl$),\linebreak ($\mbox{\it antisym-total}$) and ($trans$).
    We now argue that for every $<\ \in L(K)^N$, there is exactly one
    $\Delta \in S'$ such that $\ballot(<) \in \Delta$. Let $<\ \in
    L(K)^N$. We have $\vdash \Diamond_i \ballot_i(<)$ by
    ($ballot$). With ($\mbox{\it comp-At}$), we find that $\vdash
    \Diamond_N \ballot(<)$. Hence, $\Diamond_N \ballot(<) \in
    \Gamma_\varphi$, and there must be an mcs.\ $\Delta$ s.t.\
    $\ballot(<) \in \Delta$.  Now suppose that $\Delta' \in S'$ also
    contains $\ballot(<)$. By ($func2$), $\Delta$ and $\Delta'$
    contain the same formulae. Then $\Delta' = \Delta$, which proves
    the second part of the claim.

    As a consequence we will be allowed to use the formulae of the
    form $\ballot(<)$ as world labels in $M_\varphi$.
  
    We now prove the main claim of this proof.
    \begin{claim}
      $M_\varphi$ is a Kripke model of SCF.
    \end{claim}
    We first prove that for every mcs.\ $\Gamma$ and $\Delta$, we have
    that $R_i \Gamma\Delta$ iff for all $i \neq j$ we have that
    $\ballot_j(<) \in \Gamma$ iff $\ballot_j(<) \in \Delta$.

    First, observe that for every $i$, $R_i$ is an equivalence
    relation because by axioms ($K(i)$), ($T(i)$), ($B(i)$) and
    ($comp\cup$) all $\Box_i$ are S5 modalities.

    ($\Rightarrow$). Suppose $R_i\Gamma\Delta$. Then by definition
    $\forall \delta \in \Delta$ we have $\Diamond_i \delta \in
    \Gamma$. For any $<\ \in L(K)^N$ and $j \neq i$, suppose also that
    $\ballot_j(<) \in \Delta$. By ($exclu$), $\Box_i {\ballot_j(<)} \in
    \Delta$. Then by hypothesis $\Diamond_i\Box_i \ballot_j(<) \in
    \Gamma$, which by ($B(i)$) entails that $\ballot_j(<) \in
    \Gamma$. Because $R_i \Gamma\Delta$ is an equivalence relation,
    the same reasoning can be done to prove that if ${\ballot_j(<)} \in
    \Gamma$ then $\ballot_j(<) \in \Delta$.

    ($\Leftarrow$). Suppose $\forall j \neq i$, $\forall <\ \in L(K)^N$
    we have $\ballot_j(<) \in \Gamma$ iff $\ballot_j(<) \in \Delta$.

    Suppose that $\ballot_i(<') \in \Delta$ and $\delta \in
    \Delta$. Let us note $<_\Delta$ the preference profile $(<_1,
    \ldots <'_i \ldots <_n)$. We hence have $\ballot(<_\Delta) \land
    \delta \in \Delta$. Which by ($func2$) means that $\Box_N
    (\ballot(<) \imp \delta) \in \Delta$.

    From $(exclu$), $\Box_i \bigwedge_{j \neq i} \ballot_j(<) \in
    \Gamma$. By ($ballot$), we also have that\linebreak $\Diamond_i
    {\ballot_i(<')} \in \Gamma$. Hence, by S5, $\Diamond_i
    \ballot(<_\Delta) \in \Gamma$.

    We obtain that $\Diamond_i \delta \in \Gamma$.

    %\medskip

    We now prove that there is a linear order $<\ \in L(K)^N$ such
    that $P_i \Gamma \Delta$ iff (if $x \in V(\Gamma)$ and $y \in
    V(\Delta)$ then $x <_i y$). For every $i \in N$, we construct an
    order $<_i^{\circ}$ over the set $K^{\circ} = \{ x \in K \mid
    \Diamond_N x \in \Gamma_\varphi \}$ such that $x <_i^{\circ} y$
    iff $x \blacktriangleleft_i y \in \Gamma_\varphi$. (Note that the
    reason we restrict the preliminary construction of the preference
    order to the set of possible outcomes is because the language is
    not strong enough to talk about impossible outcomes. See Remark
    \ref{rem:lang-weak}. A careless construction could lead to a
    relation over $K$ that is not a linear order.)

    Capitalising on ($unifPref$), it is immediate that $<_i^{\circ}$
    is transitive ($4(<_i)$), antisymmetric ($antisym'$) and total and
    reflexive ($total'$). Then $<_i^{\circ}$ is a linear order over
    $K^{\circ}$.

    It is now easy to obtain a linear order $<_i$ over $K$ such that
    for all $x$ and $y$ in $K^{\circ}$ we have $x <_i y$ iff $x
    <_i^{\circ} y$.

    This completes the proof that $M_\varphi$ is a Kripke model of
    SCF.

    %%\medskip

    Then, for every consistent formula $\varphi$, there is a Kripke
    model of SCF in which $\varphi$ is satisfied.

%    $\hfill\blacksquare$
  \end{pf} 

\end{document}